\documentclass[a4paper,11pt]{article}
\usepackage{pos}

\title{Efficiency estimation of self-triggered antenna clusters for air-shower detection}
\ShortTitle{Estimation of self-trigger efficiency}

\author*[a]{P. Bezyazeekov}
\author[a]{O. Fedorov}
\author[a]{Y. Kazarina}
\author[]{O. Kopylova}
\author[b]{D. Kostunin}
\author[c]{V. Lenok}
\author[a]{S. Malakhov}

\affiliation[a]{Applied Physics Institute, Irkutsk State University, 664020 Irkutsk, Russia}
\affiliation[b]{DESY, 15738 Zeuthen, Germany}
\affiliation[c]{Karlsruhe Institute of Technology, Institute for Astroparticle Physics, D-76021 Karlsruhe, Germany}

\emailAdd{bpa@astroparticle.online}

\abstract{
Air-shower radio arrays operate in low signal-to-noise ratio conditions, which complicates the autonomous measurement of air-shower signals, i.e. without using an external trigger from optical or scintillator detectors.
A simple threshold trigger for radio detector can be efficiently applied only in radio-quiet conditions, because for other cases this trigger detects a high fraction of noise pulses.
In the present work, we study aspects of independent air-shower detection by dense antenna clusters with a complex real-time trigger system.
For choosing the optimal procedures for the real-time analysis, we study the dependence between trigger efficiency, count rate, detector hardware and geometry.
For this study, we develop a framework for testing various methods of signal detection and noise filtration for arrays with various specifications and the hardware implementation of these methods based on field programmable gate arrays.
The framework provides flexible settings for the management of station-level and cluster-level stages of detecting the signal, optimized for the hardware implementation for real-time processing.
It includes data-processing tools for the initial configuration and tests on pre-recorded data, tools for configuring the trigger architecture and tools for preliminary estimates of the trigger efficiency at given thresholds of cosmic-ray energy and air-shower pulse amplitude.
We show examples of the trigger pipeline developed with this framework and discuss the results of tests on simulated data.
}

\FullConference{37$^{\rm{th}}$ International Cosmic Ray Conference (ICRC 2021)\\
		July 12th -- 23rd, 2021\\
		Online -- Berlin, Germany}

\begin{document}
\maketitle

\section{Introduction}

For solving the challenge of independent measurements of air-shower radio emission it is necessary to apply a complex approach that includes estimations of simultaneous background context and characteristics of the received signal.
The approach also depends on detector hardware, measurements and reconstruction specificity.
Various experiments test methods of self-trigger generation, however, at the current moment there is no common approach for performing self-triggered measurements at air-shower radio arrays~\cite{arianna, anita, ovro}.

To facilitate the search of the optimal algorithms, system parameters and hardware characteristics for self-triggered radio array we decide to make a framework for developing and testing the algorithms for a generation the trigger generation by data from radio array with a focus on the future hardware implementation of these algorithms.
In the present paper we describe our approach to study and develop the procedure of self-triggering using data from the Tunka-Rex detector\cite{TunkaRex_NIM_2015}, show the structure of the developed framework and trigger generation chain and discuss the efficiency estimated for different cluster geometry. 

\section{Approach}

The main goal of our study is to define the requirements and specifications for the hardware system which receives the data from a set of antenna stations and returns trigger in case of cosmic ray event and estimate its efficiency depending on the detector geometry and primary particle parameters. 
This system needs to has a false-positive rate low enough to record most of the real cosmic-ray events.
The planned hardware implementation of the proposed trigger system is a device based on field programmable gate array (FPGA) which can be deployed at the radio array with and connected to the existing DAQ.
Unlike conventional digital chips, the logic of the FPGA operation is not determined during manufacture, but is set by programming (design), allowing you to set the desired structure of a digital device in the form of a basic electrical circuit or a program in special hardware description language.
Preliminary results of working with language Verilog show the potential possibility of implementing self-trigger algorithms on FPGA.

For finding the optimal self-triggered detector geometry we propose the dense antenna cluster approach which provides to apply separated station-level and cluster-level triggers with minimal false-positive triggering and minimal dead time.
For future experiments geometry of a single dense cluster can be easily scaled to multiply clusters by introducing a higher level trigger.

For the study of the optimal specifications of this cluster we use data from Tunka-Rex Virtual Observatory (TRVO)~\cite{TRVO}. 
For testing the trigger algorithms we tested various configurations of the cluster.
For building each configuration we use the set of Tunka-Rex events from specific stations grouped by cut according to the distance between the farthest stations.

Before testing we packed the data in a structure that mimics the hardware input from the stations cluster and contains a set of traces from given stations recorded at the same time.
Trigger pipeline also receives info about the cluster (antenna coordinates and types, additional user-defined metadata), which is used at the configuration step and for cluster-level trigger generation.

\section{Framework}
\begin{figure}[t]
	\centering
	\includegraphics[width=0.7\linewidth]{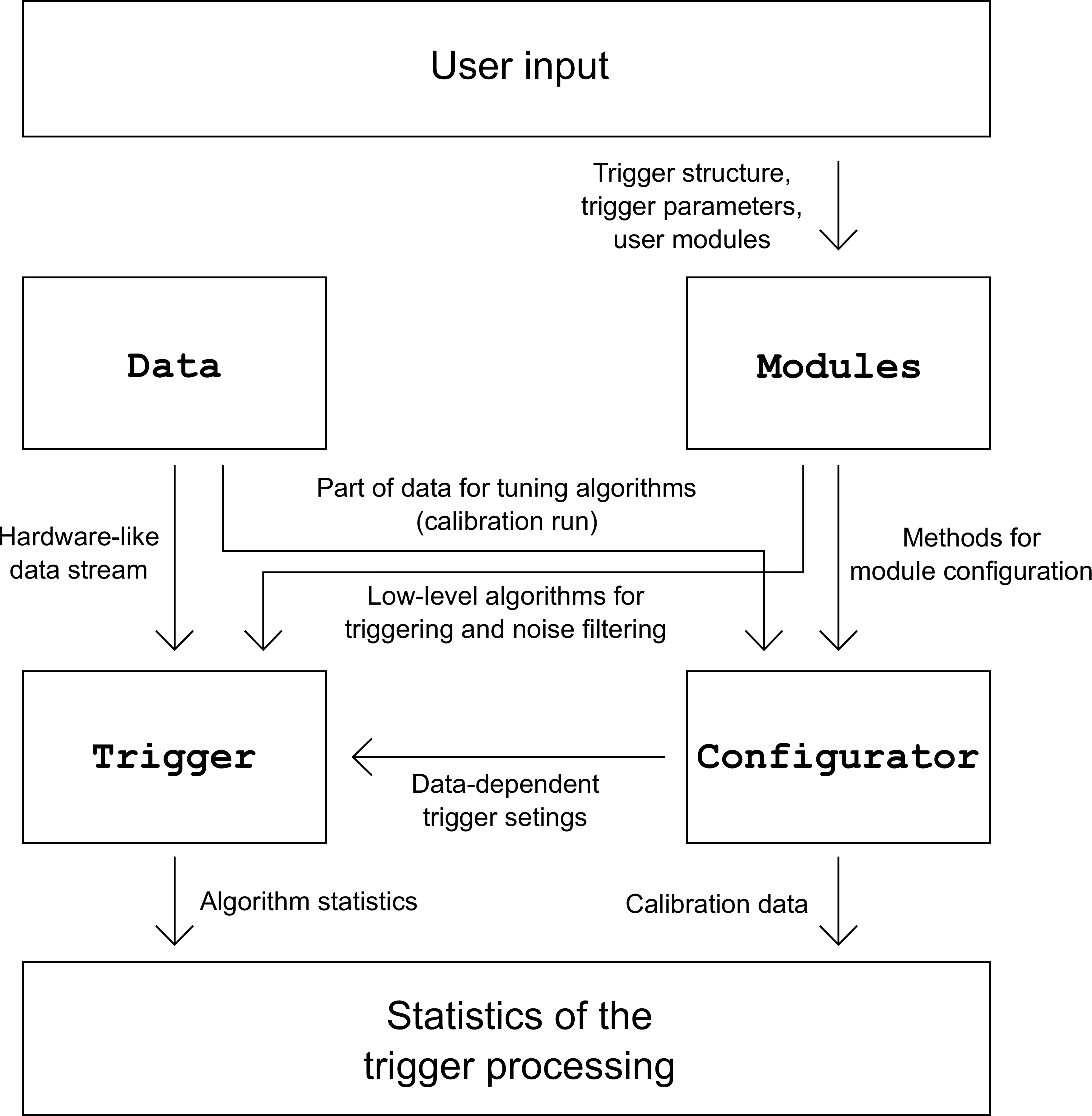}
	\\
	\caption{Structure of the framework. The framework consists of 4 main parts: \texttt{Data} is data processing tools. This part provides management and preprocessing of data. \texttt{Modules} provide the user interface for configuration of structure and parameters of trigger system and implementing user trigger or noise filtering algorithms on the Python programming language. \texttt{Configurator} is an offline part of the self-trigger. This part calculates optimal parameters of the trigger and noise suppression for antenna cluster and each station separately using data from the calibration run. \texttt{Trigger} is a software implementation of the hardware part of self-trigger. It is used for evaluation the efficiency and tuning of trigger algorithms.}
	\label{fig:scheme}
\end{figure}
The framework is designed for development and testing the methods of the trigger generation and noise suppression by using radio data from compact antenna clusters.
Compact antenna cluster in this context is conception of small-size antenna array (with distance between the farthest stations in range of few hundreds of meters) which provides to implement the approach of at least two-layer trigger generation - single-station level L0 and whole cluster level L1.
This modular system provides the possibility for flexible adjustment of the trigger generation algorithms and characteristics of the experimental equipment.
The framework includes various modules and algorithms for the trigger generation and noise suppression, also provides an interface for writing new modules and fitting to specific detector hardware.
The framework can operate in two modes:
\begin{itemize}
\item Calibration and configuration mode.
This mode is an offline part of the self-trigger algorithm. 
In this mode data from the detector is processed for defining the optimal parameters of the trigger system  according to current background conditions, requested count rate and other parameters.
We suppose that this mode will be turned on regularly depending on fluctuations of the measured signal.
\item Measurements mode.
In this mode trigger system receives detector data and processes it itself using parameters taken during the calibration.
This mode allows one to evaluate the parameters of the operation of the algorithms for generating the trigger and system in complex.
\end{itemize}
The common scheme of data flow through the trigger pipeline is shown in Fig.~\ref{fig:scheme}.
For preliminary tests framework also includes utilities for processing CoREAS~\cite{Huege:2013vt} simulations and convolution of them with generated noise.

\begin{figure}[t]
	\centering
	\includegraphics[width=1\linewidth]{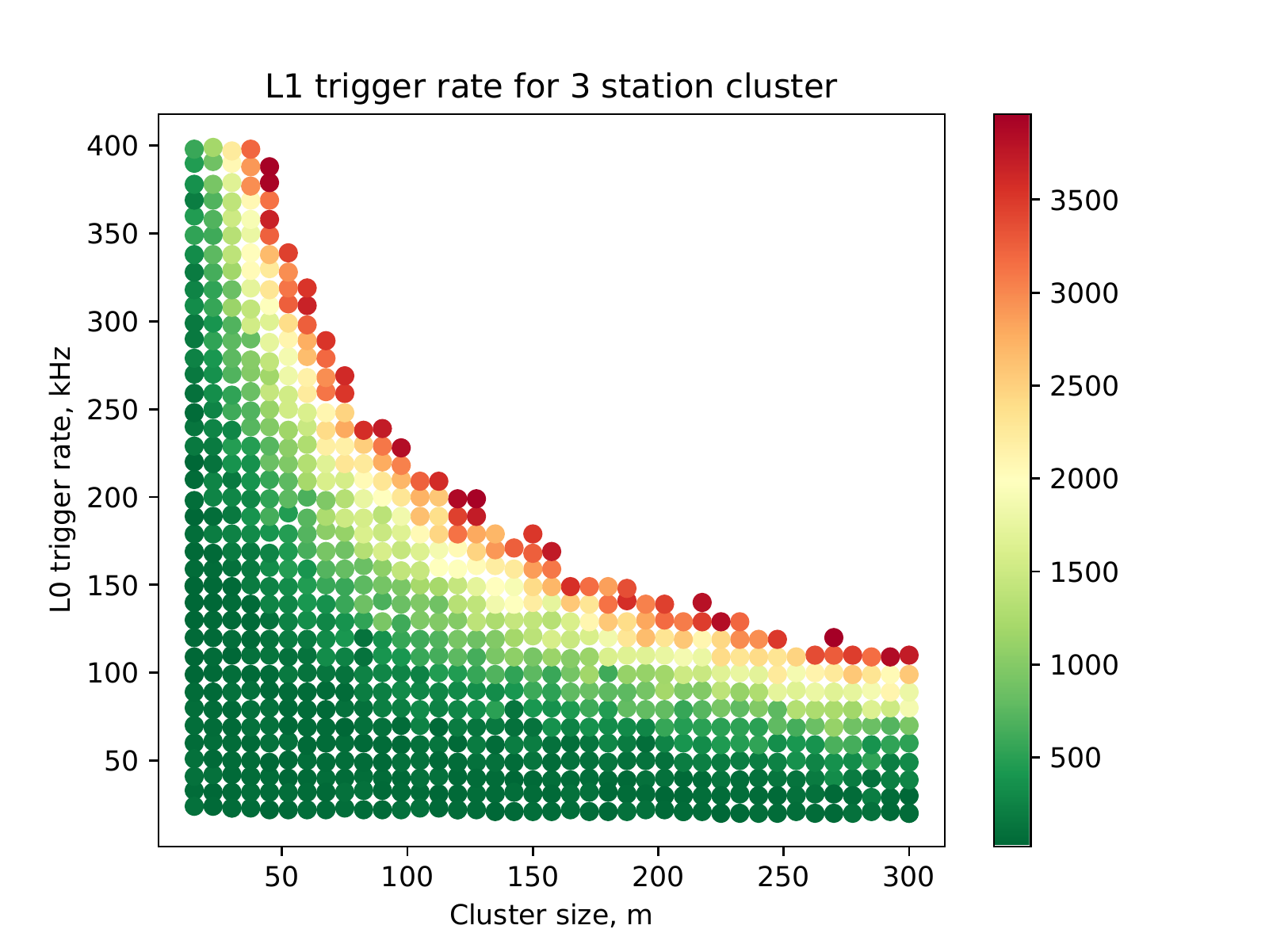}
	\\
	\caption{L1 trigger count rate depending on L0 trigger count rate and cluster size in case of 3-station cluster.}
	\label{fig:L1rate}
\end{figure}

\section{Trigger algorithm}
Trigger generation methods used by optical or scintillator detectors cannot be effectively used to generate a trigger on a radio detector.
In our approach, we use compact clusters of antenna stations and a hierarchical approach to trigger generation.
On the first level each antenna station generates a trigger independently (L0 trigger).
In the case of simultaneous (within the given time window) L0 trigger generation at several stations, second-level trigger generation algorithms are enabled (cluster level, L1 trigger).

L0 and L1 trigger generation procedures have similar architecture.
The trigger generation procedure is a chain of modules, each of them can generate a trigger from detector data or prohibit the trigger generation.

Examples of modules:
\begin{itemize}
    \item Threshold trigger. Simple trigger, which determines the probability of the presence of air-shower signal in data by an excess of amplitude threshold.
    \item RMS noise suppressor. By analysis of the 2-channel distribution of amplitudes detected on antenna this module search cores of specific RMS associated with noise pulses. 
    After defining the range of noise-related RMS, the corresponding parts of signal traces are rejected for trigger generation as noise-contaminated.
    \item Template module. This module search searches the signal by convolution with the template of averaged EAS pulse and generates the trigger.
\end{itemize}

\section{Tests on model data}

For testing the framework we choose the geometry of 3 stations triangle cluster with 200 meters spacing between stations (same as a central part of the first generation of Tunka-Rex). 
The dataset used for this study contains simulated events with energy range from $10^{17}$ to $10^{18}$ eV overlapped with real noise measured at corresponding Tunka-Rex stations.
Chain of L0 triggering consists of one module of amplitude threshold trigger, which is configured to 200 kHz count rate for given data according to limits of modern FPGA (for the relation between cluster size and L0 and L1 count rate see Fig.~\ref{fig:L1rate}).
L0 triggers are collected to the queue for the selection of the L1 triggers according to L0 timestamps.
L1 trigger chain includes arrival direction reconstructor which prohibits triggering on the pulse with zenith angle > $60^\circ$.
L1 event is defined as triggered, if all L0 triggers of this event are contained in the queue within the given time window, which depends on the size of the cluster and for this study is set to 600 nanoseconds.
Using this configuration, we estimate the fraction of triggered events from all events depending on the energy and reach the efficiency of $\approx 50\%$ for events with the energy of about 300 PeV and shower core located in a range of 200 meters from the nearby antenna.

\section{Discussion and conclusion}
Estimation of simplified trigger procedure shown principal applicability of the framework.
After the finishing we will begin to test different methods for signal and noise pulses separation.
We plan to test modules of convolution the data with air-shower pulse templates, using compressed neural networks~\cite{autoencoder} for FPGA implementation and spectral analysis. 
Test of modules will be the important step in trigger development and for this we plan to engage the astrophysical community and use additional data from other air-shower radio detectors.

\section*{Acknowledgements}
This work is supported by RSF grant № 19-72-00010.

\bibliographystyle{ICRC}
\bibliography{references}

\end{document}